\documentclass[aps,prd,groupedaddress,showpacs,nofootinbib
]{revtex4}
\usepackage{graphicx,amssymb,bm}
\usepackage{subfigure}
\usepackage{dsfont,slashed}
\usepackage{color}

\newcommand{\be}{\begin{equation}}
\newcommand{\ee}{\end{equation}}
\newcommand{\bea}{\begin{eqnarray}}
\newcommand{\eea}{\end{eqnarray}}
\newcommand{\beaa}{\begin{eqnarray*}}
\newcommand{\eeaa}{\end{eqnarray*}}

\newcommand{\nn}{\nonumber}

\def\rf#1{(\ref{#1})}

\usepackage{amsmath,amssymb}

\begin{document}

\title{Matter couplings in Ho\v{r}ava-Lifshitz and their cosmological applications}

\author{Sante Carloni$^{(a)}$\footnote{E-mail address: carloni@ieec.uab.es},
Emilio Elizalde$^{(b)}$\footnote{E-mail address: elizalde@ieec.uab.es,
 elizalde@math.mit.edu}, Pedro J. Silva$^{(c)}$\footnote{E-mail address: psilva@ifae.es}\vspace*{3mm} }

\address{$^{(a)}$  Institut d'Estudis Espacials de Catalunya
(IEEC) \\
Campus UAB, Facultat de Ci\`encies, Torre C5-Par-2a pl \\ E-08193 Bellaterra
(Barcelona) Spain\\
$^{(b)}$ Consejo Superior de Investigaciones Cient\'{\i}ficas (ICE/CSIC) \, and \\ Institut d'Estudis
Espacials de Catalunya
(IEEC) \\
Campus UAB, Facultat de Ci\`encies, Torre C5-Par-2a pl \\ E-08193 Bellaterra (Barcelona) Spain \\
$^{(c)}$ Institut de Ci\`encies de l'Espai (IEEC-CSIC) and Institut de F\'{\i}sica d'Altes Energies (IFAE)\\
UAB, E-08193 Bellaterra (Barcelona) Spain}

\pacs{04.60.Bc, 04.50.Kd, 04.60.-m, 98.80.Es, 05.45.-a}

\begin{abstract}
In this paper, the issue how to introduce matter in Ho\v{r}ava-Lifshitz theories of
gravity is addressed. This is a key point in order to complete
the proper definition of these theories and, what is very important,
to study their possible phenomenological implications.
As is well known, in Ho\v{r}ava-Lifshitz gravity the breakdown
of Lorentz invariance invalidates the usual notion of minimally coupled
matter. Two different approaches to bypass this problem are here described.
One is based on a Kaluza-Klein reinterpretation of the 3+1 decomposition
of the gravity degrees of freedom, what naturally leads to a definition of
a U(1) gauge symmetry and, hence, to a new type of minimal coupling. The
other approach relies on a midi-superspace formalism and the subsequent
parametrization of the matter stress-energy tensor in terms of deep
infrared variables. Using the last option, the phase space of the
Ho\v{r}ava-Lifshitz cosmology in the presence of general matter couplings
is studied. It is found, in particular, that the equation of state of
the effective matter may be very different from the actual matter one,
owing to the non-linear interactions which exists between matter and gravity.
\end{abstract}

\maketitle

\tolerance=5000

\section{Introduction}

Recently, Ho\v{r}ava made a proposal for an ultraviolet completion
of general relativity (GR), normally referred to as
Ho\v{r}ava-Lifshitz (HL) gravity \cite{h}, due to
Ho\v{r}ava's initial inspiration on the Lifshitz theory in solid
state physics. The salient characteristic of the HL proposal is
that it seems to be renormalizable, at least at the level of power
counting. This ultraviolet behavior is obtained by introducing
irrelevant operators that explicitly break Lorentz invariance but
ameliorate the ultraviolet divergences. On the other hand,
Lorentz invariance is expected to be recovered at low energies, as
an accidental symmetry of the theory.

The Original HL proposal has evolved in many aspects and
we count nowadays numerous sophisticated versions. In these, new
terms have been added to the original Lagrangian, with the idea to
generalize the proposal to make it more viable from the
phenomenological perspective (see \cite{Sotiriou:2009bx-gy}),  and to cure
the so-called strongly coupled problem  (\cite{strongcoupling},
\cite{Blas:2009yd} and references therein), via the introduction of
new terms \cite{Blas2}. Also, attempts to further generalize
this theory at the action level have been undertaken in \cite{Chaichian:2010yi}.  Although, as we write this article,
the consistency of the theory and its phenomenological implications remain still uncertain, it seems clear that the above extensions deserve a careful analysis.

An important feature of the original HL theory and its
modifications is the breaking of diffeomorphism invariance (Diffi) due to
the introduction of precisely those irrelevant operators that cure
the UV regime. The lower number of symmetries in the theory, as
compared to GR, produces the collateral effect that one looses the
notion of ``minimal coupling" between matter and gravity. Another
equivalent point of view (at least at the classical level) comes
from the covariant formulation of \cite{G,Blas:2009yd} where a Stuckelberg extra scalar degree of freedom over the metric field has to be
introduced that may couple to matter in many different ways.
In any case, for both formulations (the covariant and the non-covariant one)
we have in principle no arguments to choose a particular type of
coupling from amongst the most general family of couplings between the
gravity and matter sectors. There is very little work in this regard in
the literature. For example, in Ref.~\cite{Kcosmo} minimal
coupling is just assumed, to make more easy contact with GR. Other options where some particular couplings have been considered can be found, for example, in Ref.~\cite{geodesics}.

In the present paper, two different viable ways to approach this important problem are considered.
One of them is a general framework that will teach us how to
incorporate, in an educated manner, our ignorance on couplings
between matter and gravity. In fact, this method
can be used mostly in cosmology, but also in Black-Hole physics
and other general situations in which certain amount of space-time
symmetries are assumed. Here, we will import points of view and
basic methodology from midi-super-space approaches and the 3+1
decomposition. The main idea is to parameterize the total
four-dimensional energy tensor ${\cal T}_{\mu\nu}$ in terms of the deep
IR energy variables, thus
obtaining a formal expansion where the IR limit corresponds to the usual GR stress-energy tensor and higher-order terms correspond to the particular modifications introduced by the HL theory.
Since these IR variables represent well known matter that we see in
our laboratories (like for example density and pressure $(\rho,p)$),
they should satisfy the usual equation of state
and conservation laws, since the theory is assumed to recover Diffi at low energies.
Our other approach is based on a reinterpretation
of the 3+1 decomposition as a form of Kaluza-Klein dimensional reduction,
where we still have an untouched three dimensional Diffi. Then, use is made
of the fact that electro-magnetic duality in three dimensions relates
one-forms to two-forms, such that we can translate the couplings of matter with
the shift $N$, into a $U(1)$ gauge field coupling to charged
matter. At this point we recover a seance of minimal coupling based
on the gravitational $U(1)$ gauge theory. Obviously, this $U(1)$-symmetry is only relevant for the matter sector that couples to $N$ and represents, therefore, only a partial solution to the general problem.

After defining the above frameworks, we proceed to apply these ideas to cosmological scenarios.
Much research on this particular aspect of HL gravity has been done in the last two years \cite{HLcosmology}. Here, in order to address the rather involved
issue of studying the cosmological phase space of non-minimally coupled
HL gravity, we will borrow, as we already did in a previous work \cite{SES},
specific techniques from the field of dynamical systems that are
frequently used on more
canonical studies applied to diverse types of cosmologies
(see \cite{ellisbook} for a thoughtful introduction to the technique). Another example of this kind of analysis of HL gravity is given in \cite{phase-cosmo}. Later in this paper, we specifically study the {\it cosmological phase space} of the HL
model with our new matter couplings.  Our current investigation is focused
on the introduction of non-standard couplings between matter and gravity
and should be understood as a genuine extension of our previous work
on this subject \cite{SES} to cases of real phenomenological interest.

The paper is organized as follows. In Sect.~\ref{1} we present a
short overview of the relevant modification of the HL theory we
are considering, namely the inclusion of a minimal potential
defined in \cite{Sotiriou:2009bx-gy}, and the so called ``healthy
extension" of \cite{Blas2}, that presents a whole family of new
terms related to the lapse function $N$. In Sect.~\ref{2} we
present our framework to study matter couplings while in
Sect.~\ref{3} we describe an application of the above frameworks, to
cosmological scenarios with our generalized matter couplings. We
are then able to characterize the classical phase space, discussing
its structure in depth, in particular all the fixed points and
their nature as repellers or attractors in the theory with matter
couplings, in each of the corresponding cases.
Finally, in Sect.~\ref{4} we summarize the results obtained,
giving some perspectives for further work.

\section{ Ho\v{r}ava-Lifshitz without matter}
\label{1}

In HL gravity, the gravitational dynamical variables are
defined to be the lapse $N$, the shift $N_i$ and the space metric
$g_{\,ij}$, Latin indices running from 1 to 3. The space-time
metric is defined using the ADM slicing of space-time, as
\bea
ds^2=h_{\mu\nu}dx^\mu dx^\nu=-N^2dt^2+g_{ij}(dx^i+N^i)(dx^j+N^j)\;,
\eea
where $N^i=g^{ij}N_j$, as usual. The action $S$ is written in terms of
geometric objects, covariant under 3d-diffeomorphisms, characteristics of the ADM construction, like the 3d-covariant derivative $\nabla_i$, the
spatial curvature tensor $R_{ijkl}$, and the extrinsic curvature
$K_{ij}$. They are defined as follows,
\begin{equation}
R^{i}{}_{jkl}=W^i{}_{jl,k}-W^i{}_{jk,l}+ W^m{}_{jl}W^i{}_{km}-
W^n{}_{jk}W^i{}_{lm}\;,
\end{equation}
where $W^i{}_{jl}$ are the Christoffel symbols (symmetric in the lower indices), given by
\begin{equation}
W^i_{jl}=\frac{1}{2}g^{im}
\left(g_{jm,l}+g_{ml,j}-g_{jl,m}\right)\,.
\end{equation}
The Ricci tensor is obtained by contracting the {\em first} and the
{\em third} indices
\begin{equation}\label{Ricci}
R_{ij}=g^{kl}R_{ikjl}\quad\hbox{and}\quad R=g^{ij}R_{ij}\,,
\end{equation}
while the extrinsic curvature is defined as
\be
K_{ij}={1\over2N}(-\dot{g}_{ij}+\nabla_iN_j+\nabla_jN_i)\,,
\ee
where the dot stands for time derivative.

In terms of the above tensor fields, the HL action can be written as
\bea
S=\int dt\, dx^3\,N\sqrt{g}\left({\cal L}_{kinetic}-{\cal
L}_{potential}+{\cal L}_{matter}\right)\,,
\eea
being the kinetic
term universally given by \be \label{k} {\cal
L}_{kinetic}=\alpha(K_{ij}K^{ij}-\lambda K^2)\,, \ee and with $\alpha$
and $\lambda$ playing the role of coupling constants. Originally,
the potential term was a generic function of $R_{ijkl}$ and
$\nabla_i$ but in Ref.~\cite{Blas2} it was realized
that this generic function should also depend on $a_i=\nabla_i
\ln(N)$.

The action generically breaks covariance down to the subgroup of
3-dimensional diffeomorphisms and time reparametrization,
 i.e. $x\rightarrow \tilde x(t,x)$ and $t\rightarrow \tilde
t(t)$. Assigning dimension -1 to space and dimension -3 to time,
it can be seen that it is enough to restrict the potential to be
made out of operators up to dimension 6, in order to get a power-counting
renormalizable theory.

Here, we will work with a potential which corresponds to the more
general choice available, composed by the potential defined in
\cite{Sotiriou:2009bx-gy} (the SVW case), which depends on $R_{ijkl}$
and $\nabla_i$,
\bea  \label{svw}
{\cal
L}_{potential-SVW}&=&\beta_8\nabla_iR_{jk}\nabla^iR^{jk}+\beta_7R\nabla^2R
+\beta_6R^i_jR^j_kR^k_i+\beta_5R(R_{jk}R^{jk}) \\ \nonumber
&&+\beta_4R^3+\beta_3R_{jk}R^{jk}+\beta_2R^2+\beta_1R+\beta_0\,,
\eea
and with the addition of all the general terms as suggested in \cite{Blas2}.
We collect all these terms in an implicit form
\be {\cal L}_{potential-a_i}=\sum
\gamma_n\,O^n(a_i,\nabla_j,R_{ijkl}), \label{bps}\ee
where $O^n$ are general operators of maximum dimension 6 and
$\gamma_n$ the corresponding coupling constants.

In our calculation, we have worked out all the independent terms of
these operators and have chosen to display only, as the representative operator
for each class, the ones with less derivatives acting on a
single $a_i$\footnote{This is an arbitrary basis that,
nevertheless, fixes our conventions.}. We therefore obtain:
\bea
\hbox{At order}\, &2& \nn \\&&{\gamma_0\, R}+ {\gamma_1\, a^2}\,,
\label{Ord2} \\
\hbox{at order}\, &4&\nn\\ &&\gamma_3 \,a^4+ \gamma_4\, a^2D_i a^i
+ \gamma_5\, D_i a^i D_j a^j+{\gamma_6\, a^2 R}+{\gamma_7\, a^ia^j
R_{ij}}+{\gamma_8\,D^2 R}+{\gamma_9\, R_{ij}
R^{ij}}+{\gamma_{10}\,R^2}\,,\label{Ord4}\\
\hbox{and at order}\, &6& \nn\\ &&\gamma_{11}\, a^6+{ \gamma_{12}\,
a^4R}+{ \gamma_{13}\, a^2 a_i a_j R^{ij}}+{\gamma_{14}\, a^2
R^2}+{\gamma_{15}\, a_ia_j R^{ij}R}
+{\gamma_{16}\,a_ia_k R^{ij}R^{k}_{j}}+{\gamma_{17}\, R^3}\nn\\
&&+{\gamma_{18}\,R_{ki}R^{ij}R^{k}_{j}}+{\gamma_{19}\,RR^{ij}R_{ij}}+{\gamma_{20}\,a^2R^{ij}R_{ij}}
+\gamma_{21}\,a^4D_i a^i +\gamma_{22}\,a^2D_i a^i D_j a^j+\gamma_{23}\,D_i a^i D_j a^jD_k a^k\nn\\
&&+\gamma_{24}\,D_i a^j D_j a^iD_k a^k+\gamma_{25}\,D_i a^j D_k
a^iD^k a_j+{\gamma_{26\,}a^2 a^kD_k R}+{\gamma_{27}\,a^i a^kD_k
D_iR}+{\gamma_{28}\,a^2 D^2 R}\nn\\
&&+{\gamma_{29}\,D^2 D^2R}+{\gamma_{30}\,a^ia^j R_{ij}D_ka^k}+{\gamma_{31}\,a_iD R^{i}D_ka^k}
+\gamma_{32}\,a^jR_{ik}D^iR^{k}_{j}+{\gamma_{33}\,D^jR_{ik}D^iR^{k}_{j}}\nn\\
&&+{\gamma_{34}\,RD^2R}+{\gamma_{35}\,D^iRD_iR}
+{\gamma_{36}R^{ij}D_iD_jR}\,. \label{Ord6} \eea
The above list includes previous terms of the SW-potential and a number of new terms due to the appearance of the new field $a$. Notice that we have defined the potentials such that they have to be multiplied by $N\sqrt{g}$, which explicitly contains $N$. Therefore, even though some of the new terms in the potential do not exhibit
an explicit coupling with $a_i$, they are {\it not} equivalent to any
term of the SVW-potential.

At this point, other phenomenological and theoretical
considerations may help us constraint the range of values the
different couplings should take. For example, in
Ref.~\cite{Bogdanos:2009uj} it was found that ghost instabilities are
present if $\lambda\in (1/3,1)$, that the cosmological constant is
negative for the detailed balance potential, $\alpha > 0$, and so on.
Here, we will constraint as little as possible the different
ranges of values on each coupling constant to see how much
information comes out of the dynamical system approach itself. Then, we
will add this information to the constraints arising from other
considerations, to finally obtain the most promising form of the
potential. In particular, we will take $\lambda$ different from
1/3 (corresponding to the scale invariant case) as the only
limitation on its range.

\section{Mater couplings}
\label{2}

Once we have explicitly defined the extension of the
HL theory we will be working with---at least what concerns its gravity sector---it is time  to focus now on how matter is to be coupled to
gravity. As we mention in the introduction, due to the reduction
of the symmetries present in the theory, we have no longer a valid
argument to define a minimal coupling. In fact, this is more
dangerous than what it may naively seem, since for example, different
particles will have in general different dispersion relations,
depending on their couplings with the gravity sector; also, since we have more geometric invariants, there are many more ways to construct couplings to matter. We
postpone this line of thought to future work, to focuss here just on the
the simplest problem of parameterizing the possible form of all
these different couplings in terms of physical quantities.

\subsection{Midi-superspace approach}

The first approach is based on two main assumptions. First, that in the deep IR regime one should recover diffeomorphism invariance and hence, that the corresponding IR stress energy tensor $T_{\mu\,\nu}$ should be divergence-free. Using the above, we can formally expand the full stress energy tensor, $\cal T_{\mu\nu}$, in terms of physical observables, defined in the deep IR regime alone. Second, we assume that we have at disposal a considerable number of symmetries, like in cosmological models or Black Hole physics, which allow us to write $\cal T_{\mu\,\nu}$ in terms of just a few variables, as the energy density $\rho$, the pressure $p$, fluid velocity $v$, etc. The above assumptions imply that we can write
\bea
{\cal T}_{\mu\nu}=T_{\mu\nu}(\rho,p,\ldots)+\Delta T_{\mu\nu}(\rho,p,\ldots),
\eea
where $\Delta T_{\mu\nu}(\rho,p,\ldots)$ is the leftover contribution, made out of irrelevant operators and other such terms which will anyway decouple at low energies. The above expansion can be understood as a sort of derivative expansion in the gravitational coupling with matter fields. From the point of view of the covariant formulation, where a Stuckelberg field $\phi$ is added on top of the metric $g$, what we are here doing is to separate the matter Lagrangian ${\cal L}_{matter}(\psi)$, where $\psi$ represents matter fields, into a part which is minimally coupled ${\cal L}_{min}(g;\psi)$ and the rest of it, ${\cal L}_{non-min}(\phi,g;\psi)$, which generically has couplings to $\phi$ and $g$, e.g.,
\be
{\cal L}_{matter}(\psi)={\cal L}_{min}(\psi;g)+{\cal L}_{non-min}(\psi,g;\phi).
\ee

Using the above ideas, we can describe our HL theory in terms of a Lagrangian formulation where the gravity sector is given in terms of midi-superspace variables, while the matter sector is described in terms of hydrodynamic variables, like $(\rho,p)$. For example, consider the case of the celebrated FRW ansatz. Here, due to the symmetries imposed, ${\cal T}_{\mu\nu}$ depends on two variables $(\rho,p)$ only and it can be written so that it is diagonal. Then, our previous considerations translate into the following expansion
\bea
{\cal T}_{00}=\rho^T=\rho+\delta_0p+\delta_1p^2+\delta_2Rp+\delta_3Rp^2+\delta_4R^2p+
\delta_5p^3+\delta_6\rho,\label{couplingRho}\\
\frac{g^{ij}{\cal T}_{ij}}{3}=p^T=p+\eta_0p+\eta_1p^2+\eta_2Rp+\eta_3Rp^2+\eta_4R^2p+\delta_5p^3+\eta_6\rho,
\label{couplingP}
\eea
where, for consistency with the gravity sector, we have limited the expansion to operators up to order 6. Note that here $(\rho, p)$ represent standard matter and, therefore, they fulfill the usual linear relations
\be
p=w\rho\,.
\label{eqnstate}
\ee
Moreover, owing to the non-minimal couplings to gravity, the above equation of state gives rise to non-linear relations for the effective total energy density and pressure $(\rho^T,p^T)$.
These equations, together with the equation of state, will define the type of fluid we can consider in our HL cosmology where matter is coupled to gravity in the most generic form.

Coming back to the general case, as in GR, minimization of the action $S$, upon variation of the metric in the pure gravity sector, defines the two-index tensor $H\!\!L_{\mu\nu}$,
\be
H\!\!L_{\mu\nu}=\frac{\delta }{\sqrt{-h}\,\delta h^{\mu\nu}}\left(\int dt\, dx^3\,N\sqrt{g}\left({\cal L}_{kinetic}-{\cal
L}_{potential}\right)\right)\,.
\ee
This tensor can also be decomposed into an IR part, corresponding precisely to the Einstein tensor $G_{\mu\nu}$, plus a leftover, characteristic of the HL theory, which we write as
\be
H\!\!L_{\mu\nu}={1\over 2\kappa^2}\left(G_{\mu\nu}+\Delta H_{\mu\nu}\right)\,,
\ee
where $\kappa^2$ is the gravitational coupling constant (in natural units $\kappa^2=8\pi G_N$). Therefore, the form of the field equations obtained by minimizing the HL action with respect to the metric is
\be
G_{\mu\nu}+\Delta H\!\!L_{\mu\nu}=\kappa^2\left(T_{\mu\nu}+\Delta T_{\mu\nu} \right)\,.
\label{hleqn}
\ee
At this point we still have to add a last constraint, that comes from taking the 4d divergence to the gravitational field equations \rf{hleqn} and using the usual Bianchi identities for $G_{\mu\,\nu}$ and $T_{\mu\,\nu}$. This yields namely
\be
\left._4\nabla\right.^\mu(\Delta H\!\!L_{\mu\,\nu})=\left._4\nabla\right.^\mu(\Delta T_{\mu\,\nu})\,,
\label{biancci}
\ee
where $\left._4\nabla\right.$ is the 4d covariant derivative. It is important to recall that these equations have to be satisfied only on-shell, since they actually come from the field equations.

Summarizing, after the whole derivation has been carried out, our final set of equations, that define our recipe to deal with general couplings of matter to HL gravity, is given by Eqs.~(\ref{eqnstate},\ref{hleqn},\ref{biancci}). Observe that the above expressions are written in terms of a set of operators of order less or equal than 6, and are made out of geometric objects, like the 3d curvature $R$, the vector $a_i$, and the 3d covariant derivative $\nabla$, and of IR hydrodynamic variables, as for example ($\rho$, $p$).

\subsection{U(1) gravitational coupling}

As we have argued previously, due to the break down of diffeomorphism invariance, there is no more a clean argument to define what it will be a minimal coupling between gravity and the matter sectors. One could still try, of course, to use a principle based merely on simplicity, but even then, it is a fact that such a principle will by no means be universal, not to talk on its grounding from pure physical considerations. Instead of following the above line of thought, we will here argue that the remaining symmetries of the theory are still strong enough in order to help us find a clear guiding principle which, under some general assumptions, will deliver a well defined definition of what a minimal coupling in HL theories of gravity should be.



To illustrate our idea, let us consider the more relevant part of the potential terms in the gravity sector of the HL Lagrangian,
\be
\int{dtdx^3\,N\sqrt{g}\left(\alpha_0R+\alpha_1a^ia_i\right)}.
\ee
We can always device a conformal transformation in the 3d metric, of the form $g_{ij}=N^{-2}\tilde{g}_{ij}$, such that, in the new frame, the vector $a$ and the 3d curvature tensor $\tilde{R}$ are canonically normalized with respect to the 3d metric $\tilde{g}_{ij}$, i.e.
\be
\int{dtdx^3\sqrt{\tilde{g}}\left(\alpha_0\tilde{R}+\gamma\tilde{g}^{ij}a_ia_j\right)},
\ee
where $\gamma=(\alpha_1-2\alpha_0)$ and where we have discarded pure boundary terms for simplicity. In this new frame, $a_i$ is a 3d one-form, which can be transformed into a two-form $F_{jk}$ via Hodge-duality. The resulting two-form is naturally described in terms of a one-form gauge potential $b_i$, what immediately leads to a gravitational U(1) gauge symmetry:
\be
a^i={1\over \sqrt{2^{3}}}\varepsilon^{ijk}F_{jk}\,,\qquad F_{ij}=\partial_i b_j-\partial_j b_i,
\ee
where $\varepsilon$ is the 3d Levi-Civita pseudo-tensor in the new frame. The constant factor in the above definition is chosen to have a canonically normalized kinetic term in the corresponding action,
\be
\int{dtdx^3\sqrt{\tilde{g}}\left(\alpha_0\tilde{R}+{1\over 4g^2}\tilde{g}^{ij}\tilde{g}^{kl}F_{ik}F_{jl}\right)},
\ee
where $g^{-2}=\gamma$.

From this point of view, we have a local $U(1)$ symmetry that may be used to define the meaning of our new ``minimal coupling" with this gravity sector. Notice that, in order to respect such symmetry, we can only write gauge invariant operators in our matter Lagrangian and, therefore, it is clear that our minimal coupling should be written in terms of covariant $U(1)$ derivatives, of the form
\be
D_i=\nabla_i-ib_i\,.
\ee
For example, the Lagrangian for a charged scalar field, $\psi$, minimally coupled to the gravity sector, will be given by
\be
{\mathcal L}=(\sqrt{g})\left[{1\over N^4}\partial_t \psi\partial_t \psi^* - g^{ij}D_i\psi D_j\psi^*-V(\psi^2)\right],
\ee
where, as a complementary principle, we have used the fact that, at low energy, we should necessarily recover the usual meaning of minimal coupling.

The case of vector fields and their possible couplings to gravity is more subtle.  A gauge field $A$ can be coupled to a conserved current $J$ in the minimal form as $J^iA_i$. In 3d we have a topologically conserved charge, made out of our gravitational gauge field $b$, namely $J^i=\varepsilon^{ijk}F_{jk}$. This current is conserved by the corresponding 3d Bianchi identities and, therefore, opens the possibility of including such a term in the usual gauge invariant action, producing thereby a sort of Chern-Simons term coupling the matter gauge field to the gravity gauge field.

The above idea is far from having been investigated in detail and deserves a lot more attention. We believe that it is indeed an interesting starting point to study such a complicated subject as is the coupling of matter to HL gravity. Note that, as an outcome of these ideas, we are left with a sort of mixed argument, based on symmetries and the low energy emergence of 4d diffeomorphism invariance, which still does not have a unique, well defined, general meaning as a ``minimal coupling" in HL gravity. Nevertheless, we have clearly achieved some relevant improvement, since at least the gravity sector related to all the operators made out of $a_i$ is under control, due to the gravitational $U(1)$ symmetry. It will be very interesting to study the implications of the above ideas to BH physics, where the form of the metric clearly induces the coupling with the $a_i$ vector. Hopefully these studies will appear soon elsewhere. In what follows here, we will directly concentrate on cosmological scenarios where, owing to the form of the typical ansatz, the $U(1)$ gravitational symmetry is trivial and we still do not take real advantage of the above mechanism.

\section{Non-minimal Cosmology}
\label{3}

In this section we apply the above methods to study the impact of non-minimal couplings\footnote{We use this term, in contraposition to the usual minimal coupling in GR, to denote a general coupling of matter to HL gravity.} to HL gravity in cosmological scenarios. We start, as our basic ansatz, from the usual FRW metric corresponding to homogeneous and isotropic space-time, i.e.
\be
ds^2=-dt^2+a(t)^2\gamma_{ij}dx^idx^j,
\ee
where $\gamma_{ij}$ is a maximally symmetric metric of constant curvature $k=(-1,0,1)$, and $a$ is the scale factor such that
\begin{equation}
R_{ij}=\frac{k}{a^2}g_{ij}\,,\qquad K_{ij}= -Hg_{ij}\,,\qquad H=\frac{\dot{a}}{a}\,.
\end{equation}
With these symmetries, the total stress energy tensor ${\cal T}_{\mu\nu}$ can be written as
\begin{equation}
{\cal T}=\rho^Tdt\otimes dt+p^Tg_{ij}dx^i\otimes dx^j,
\end{equation}
where  $\rho^T$ and $p^T$ are the effective total energy density and effective total pressure, which can be expanded as in Eq.~(\ref{couplingRho}-\ref{couplingP}).

The gravitational sector is drastically simplified in the FRW ansatz where only the terms corresponding to the coupling constant $\beta_0,\beta_1,\beta_2,\beta_3,\beta_4,\beta_5,$ and $\beta_6 $ in Eq.~\rf{svw} are not identically zero. In particular, this means that all terms related to the field $\partial \ln(N)$ do not contribute. Following, the work of \cite{Sotiriou:2009bx-gy}, we assume that our coordinates are such that $c=1$. We have also used time re-parametrization plus three-dimensional diffeomorphism invariance, in order to eliminate redundant degrees of freedom.

In this setting, the non-zero components of the HL tensor are
\begin{eqnarray}\label{HLtensor}
&&H\!\!L_{00}=\alpha\left[3\left(1-\frac{3 \xi }{2}\right)H^2+\frac{3 k}{ a^2}-\Lambda-\frac{\chi _3 k^2}{2 a^4}-\frac{\chi_4 k}{2 a^6}\right] \label{HL00}\;,\\
&&H\!\!L_{ij}=-\alpha g_{ij}\left\{ \left(1-\frac{3 \xi }{2}\right)\left[ 2\dot{H}+3H^2\right]+\frac{k}{a^2}-\Lambda+\frac{\chi _3 k^2}{6 a^4}+
\frac{\chi _4 k}{2 a^6}\right\}\label{HLij}\;,
\end{eqnarray}
where we set
\begin{eqnarray}
&16\pi G_N=\frac{1}{\alpha}\;,\qquad \Lambda=\frac{\beta_0 \alpha^3}{2}\;,\qquad 1=-\beta_1\alpha^2\;,\nonumber \\
&\chi_3=12\alpha\left(3 \beta_2+\beta_3\right)\alpha^2\;,\qquad
\chi_4=24\left(9 \beta_4+3 \beta_5+\beta_6\right)\alpha^4\;.
\end{eqnarray}
Therefore, we have
\bea
&&\Delta H\!\!L_{00}=-\left({9\over2}H^2+\frac{\chi_3 k^2 }{2a^4}+\frac{\chi_4 k }{2a^6}\right), \\
&&\Delta H\!\!L_{ij}=g_{ij}\left[ \frac{3}{2}\xi\left( 2\dot{H}+3H^2\right)-\frac{\chi _3 k^2}{6 a^4}-\frac{\chi _4 k}{2 a^6}\right],
\eea
while the constraint equation (\ref{biancci}) in this framework can be written as
\be
\dot{(\Delta H\!\!L_{00}})+3H\left(\Delta H\!\!L_{00}+{1\over 3}\Delta H\!\!L^i _{\;i}\right)=\dot{(\Delta {\cal T}_{00})}+3H\left(\Delta{\cal T}_{00}+{1\over 3}\Delta{\cal T}^i _{\;i}\right)\,.
\ee
It is possible to understand  this equation as a measure of up to which level the failure of diffeomorphism invariance in the gravity sector is transmitted to the matter sector. Nevertheless, for the FRW ansatz, it is not difficult to verify that $\left._4\nabla\right.^\mu(\Delta H\!\!L_{\mu\,\nu})=0$, due to the homogeneity of the ansatz and, therefore, we are left with the following pure constraint on the matter sector:
\be
\dot{(\Delta {\cal T}_{00})}+3H\left(\Delta{\cal T}_{00}+{1\over 3}{\Delta\cal T}^i _{\;i}\right)=0.
\ee
This non-trivial relation for the matter sector ultimately reduces the number of independent couplings that characterize the expansion (\ref{couplingRho}-\ref{couplingP}). In fact, one finds the following relation between these couplings:
\bea
\delta_6+\delta_0w=\eta_0+\eta_6w\,,\qquad\delta_1(1+2w)=\eta_1\,,\nonumber\\
\delta_5(2+3w)=\eta_5\,,\qquad \delta_2(w+2/3)=\eta_2\,,\nonumber\\
\delta_3(5/3+2w)=\eta_3\,,\qquad \delta_4(w+4/3)=\eta_4\,,
\label{w-general}\eea
for general values of $w$.

The field equations for the gravity sector, $H\!\!L_{\mu\nu}={1\over 2}{\cal T}_{\mu\nu}$ result into the following expression
\begin{eqnarray}\label{FriedNoDB}
&\left(1-\frac{3 \xi }{2}\right)H^2-\frac{\chi _2  k}{6 a^2}-\frac{\chi _1}{6}-\frac{\chi _3 k^2}{6 a^4}-\frac{\chi_4 k}{6 a^6} -\kappa ^2
\rho^T=0
\;,
\end{eqnarray}
which, together with Eqs.~(\ref{eqnstate}), (\ref{couplingRho}-\ref{couplingP}),  and (\ref{w-general}) define our cosmological system. The presence of non minimal couplings of  the matter terms conveys the idea that, because of the Lorentz violation, what drives here the cosmological expansion is no more $\rho$, but actually $\rho^{T}$. As a consequence, depending on the values of the parameters $\delta$, a certain type of matter can rather behave effectively as a fluid with different thermodynamical properties. The natural question being then, wether such behavior can possibly help explaining some important features of the observed Universe. In order to answer it, let us first investigate Eq.~\rf{FriedNoDB} for the two basic classical matter types: dust and radiation.

\subsection{The dust case ($w=0$)}
\label{3.1}
In the case of dust,  the last equation  above reduces to
\begin{eqnarray}
&\left(1-\frac{3 \xi }{2}\right)H^2-\frac{\chi _2  k}{6 a^2}-\frac{\chi _1}{6}-\frac{\chi _3 k^2}{6 a^4}-\frac{\chi_4 k}{6 a^6} -\kappa ^2
(1+\delta_6)\rho=0\label{FriedNoDB2}
\;,
\end{eqnarray}
which is exactly the same system treated in \cite{SES} with the only difference that the coupling constant is now modified by the parameter $\delta_6$. That is to say, we now have control on the way in which matter couples to gravity at the cosmological level. Note however that such small change can induce a great deal of difference in the behavior of the respective cosmology. For example, in the case $\delta_6<-1$, $\rho^T$ has a negative coupling constant, which generates an effective dark energy, even if the standard matter has the usual thermodynamical properties. In standard GR this would lead to an irreparable inconsistency of the theory. However, in HL gravity the presence of the additional terms does {\it not} exclude such case. Of course, once the coupling constant is set to be negative it stays so forever, and its effect on the cosmic processes typical of the dust era should be investigated carefully.

\subsection{The radiation case ($w=\frac{1}{3}$)}
\label{3.2}
Let us consider now the case of radiation.   The cosmological equations read

\begin{eqnarray}
&&H^2 \left(1-\frac{3 \xi}{2}\right)-\frac{\delta_5 \rho _0^3}{27 a^{12}}-\frac{2 k \delta _3
   \text{$\rho $0}^2}{3 a^{10}}-\frac{\beta_1}{9 a^8}-\frac{ \beta_2}{6 a^6}-\frac{\beta_3}{6 a^4}-\frac{k \chi _2}{6 a^2}-\frac{\chi _1}{6}=0,\label{FriedRad}\\
   &&\left(\frac{3 \xi }{2}-1\right) H'+
   \left(\frac{3 \xi }{2}-1\right)H^2-\frac{5 \delta _5 \rho _0^3}{27
   a^{12}}-\frac{8 k \delta _3 \rho _0^2}{3 a^{10}}-\frac{\beta_1}{3 a^8}-\frac{\beta_2}{a^6}-\frac{\beta_3}{6 a^4}+\frac{\chi _1}{6}=0\label{RayRad},
\end{eqnarray}
being
\begin{eqnarray}
&&\beta_1=\rho _0 \left(12^2 k^2 \delta _4+\delta _1
   \rho _0\right),\\
&&\beta_2=k^3 \chi _4+12 k \delta _2 \rho
   _0,\\
&& \beta_3=k^2 \chi _3+2 \left(\delta _0+3 \delta
   _6+3\right) \rho _0.
\end{eqnarray}
Differently from the previous situation, here the non-vanishing pressure ``switches on''  new terms associated to the additional matter couplings. Looking at the general structure of the equations above, however,  it is clear that the effects of these couplings will only be relevant at early times, namely when the scale factor $a$ is particularly small. Thus, we can conclude that the introduction of a full matter coupling will influence the evolution of the early Universe, mainly.

It is also interesting to note that, when radiation dominates, the presence of this coupling induces differences between the HL cosmology and the GR one, even in the case of spatially flat solutions. Such difference is not present in the dust case.

In order to have a more clear idea of the effects of the matter coupling in Eqs.~(\ref{FriedRad}-\ref{RayRad}) we can use the dynamical system approach. Given the high number of degrees of freedom of the system, we will limit ourselves to consider the finite analysis only.
Following the method of \cite{SES}, we define the variables
\begin{eqnarray} \label{}
\nn&&X=\frac{\chi _1}{3 (3 \xi -2) H^2},\quad Y=\frac{2 \delta _5 \rho_0^3}{27 (3 \xi -2) H^2 a^{12}},\quad Z=\frac{4 k \delta _3\rho_0^2}{3 (3 \xi -2) H^2 a^{10} },\\ &&R=\frac{\beta _1}{9 (3 \xi -2)
   H^2 a^8},\quad S=\frac{\beta _2}{3 (3 \xi -2) H^2  a^6},\quad T=\frac{\beta _3}{3 (3
   \xi -2)  H^2 a^4},\\ \nn&& K=\frac{k \chi _2}{3 (3 \xi -2) H^2 a^2 },
\end{eqnarray}
through which we will characterize the phase space, and a logarithmic time $N=\ln a$.

Then, the resulting dynamical system is
\begin{eqnarray}
\nn &&X'=-2 X (3 R+2 S+T-X+5 Y+4 Z-1),
\\  \nn &&Y'=-2 Y (3 R+2 S+T-X+5 Y+4 Z+5),
\\ &&Z'=-2 Z (3 R+2 S+T-X+5 Y+4 Z+4), \label{DynSysRad}
\\  \nn &&R'=-2 R (3 R+2 S+T-X+5 Y+4 Z+3),
\\ \nn && S'=-2 S (3 R+2S+T-X+5 Y+4 Z+2),
\\ \nn &&T'=-2 T (3 R+2 S+T-X+5 Y+4 Z+1),
\\ && 0 =1 + K + R + S + T + X + Y + Z,
\end{eqnarray}
where the ``prime'' indicates derivative with respect to $N$. The structure of the system reveals that the phase space is divided into different sectors, delimited by invariant submanifolds, and that, as a consequence, no global attractor can actually exist\footnote{To be precise, this role could be taken by a point at the origin. However, as we will see, such point would be always unstable.}. Therefore, any orbit that can have physical interest will be realized by only using a restricted set of initial conditions.

The fixed points can be found, as usual, by setting the lhs of the equations equal to zero. Then, the solutions associated to the fixed points can be found using the general expressions
\begin{eqnarray}
&& \dot{H}=\alpha H^2 , \label{eq:dyn_sys_a} \\\
&&\frac{ \dot{\rho}}{\rho}\, = 4 H= \, - \frac{3}{\alpha (t-t_0)} ,
	\label{eq:dyn_sys_b}\\
&& \nn \alpha = -1 + 3 R_i + 2 S_i + T_i - X_i + 5 Y_i + 4 Z_i, \,
\end{eqnarray}
where  the ``i'' subscript represents the value of the corresponding variable at the fixed point. Subsequently, using the Hartman-Gro\ss mann theorem, we can investigate their stability. The finite fixed points, their associated solution and their corresponding stability are all summarized in Table \ref{FP finite}.

As expected, we find fixed points associated to the dominance of the different terms in Eqs.~(\ref{FriedRad}-\ref{RayRad}). These points are characterized by the corresponding expansion rates of the scale factor and of the dissipation of the energy density. Choosing initial conditions in which all the variables are negative (which correspond to a specific set of constraint for the coupling constants $\chi_i$ and $\delta_i$), we will select the sector of the phase space in which the orbits have access to all fixed points. In the same way, cosmic histories that surely avoid any of  the states described by one of such points can be selected by modifying the values of the couplings.

Note also that none of the fixed points related to the matter couplings appears to be characterized by an accelerated expansion, regardless of the relative sign of the coefficients of those terms. We will discuss in the conclusions the possible consequences of the presence and nature of these points.
And note moreover that, in spite of all these changes one still has a de Sitter attractor that is generated by the cosmological terms in the gravitational part of the action.  Therefore, the introduction of the matter couplings does not, in principle, compromise the formation of a dark energy era, in spite of the nature of the cosmic fluid and of the structure of the non-minimal couplings. This is a quite remarkable result.

Substitution in the cosmological equations reveals that the fixed points $\mathcal{A}$, $\mathcal{C}-\mathcal{G}$  are not actual solutions of the cosmological equations. However, this is not a serious problem, since all these points are unstable and, therefore, they may only correspond to an approximation of the general solution represented by the full orbits. On the contrary,
the point $\mathcal{B}$ corresponds, instead, to an exact solution of the theory.

\begin{table}[htdp]
\caption{Coordinates, solutions, and stability of all the fixed points of the dynamical system \rf{DynSysRad}.}
\begin{center}
\begin{tabular}{clllllllllllll}\hline
Point&~~~&Coordinates&~~~& Solution&~~~& Energy Density&~~~& k &~~~&Stability\\
& ~~&[X,Y,Z,R,S,T]& & \\\hline\\
$\mathcal{A}$&& $[0,0,0,0,0,0]$ & &$a=a_0 (t-t_0)$& &$\rho\, = \, (t-t_0)^{-4}$& &-1 &&saddle\\ \\
$\mathcal{B}$& &$[-1,0,0,0,0,0]$& &$a=a_0 \exp\left(\sqrt{\frac{\chi1}{3(2-3\xi)}}\; t\right)$& &$\rho\, = 0 $&& 0&&attractor\\ \\
$\mathcal{C}$& &$[0,-1,0,0,0,0]$ & &$a=a_0 (t-t_0)^{1/6}$& &$\rho\, = \, (t-t_0)^{-2/3}$& &0 & &repeller\\ \\
$\mathcal{D}$& &$[0,0,-1,0,0,0]$ & &$a=a_0  (t-t_0)^{1/5}$& &$\rho\, = \, (t-t_0)^{-4/5}$& & 0 & &saddle\\ \\
$\mathcal{E}$& &$[0,0,0,-1,0,0]$ & &$a=a_0  (t-t_0)^{1/4}$& &$\rho\, = \, (t-t_0)^{-1}$& &0 & &saddle\\ \\
$\mathcal{F}$& &$[0,0,0,0,-1,0]$  &&$a=a_0  (t-t_0)^{1/3}$& &$\rho\, = \, (t-t_0)^{-4/3}$&& 0&&saddle\\ \\
$\mathcal{G}$& &$[0,0,0,0,0,-1]$ & &$a=a_0  (t-t_0)^{1/2}$& &$\rho\, = \, (t-t_0)^{-2}$&& 0&&saddle\\
\\ \hline
\end{tabular}
\end{center}
\label{FP finite}
\end{table}%

\section{Discussion and conclusions}
\label{4}

We have proposed in this paper two different ways to study the coupling of HL theories of gravity with matter. First, we have devised a very natural procedure to study the effect of matter couplings in a Ho\v{r}ava-Lifshitz theory of gravity, based on the imposition of the Bianchi identities on a midi-superspace approach together with the assumption of a well-behaved IR behavior. The use of these geometric relations has revealed that, when matter is present, we need to supplement the theory with an additional constraint that ensures the consistency of the gravitational and matter sectors.  
Second, we have introduced an alternative definition of  `minimal coupling' based on the $U(1)$ gauge symmetry present in a very important sector of the theory. In this way, we are able to couple the matter fields via a gauge invariant formalism. These couplings are both relevant for BH theories and also for perturbation theory in a cosmological scenario, but irrelevant for the study of cosmological solutions, since in this last case the corresponding gravity sector is identically zero.

Regarding the application of the above ideas to cosmological scenarios, in order to put a first set of constraints on this new version of the theory we have chosen to analyze its Friedmannian cosmology. The high symmetry of cosmological space-times greatly simplifies the equations and allowed us to introduce, in a simple way, a set of (arbitrary) additional matter couplings. In particular, the constraint simplifies into algebraic relations among the coupling constants of the additional matter terms, which can be used to understand their specific physical role.

A first interesting finding here was that, because of the nature of the constraint, the behavior of the HL Universe is very sensitive to the thermodynamical properties of matter.  The cosmological equations are enriched with additional terms and both their structure and their number change with the barotropic factor. This implies that the resulting phenomenology is expected to be quite different from one case to another, when one considers Universes dominated by different types of matter.

In addition, although in principle the new matter couplings influence the entire cosmic history, from the structure of the equations it was easy to conclude that the new terms have more weight at early times. This suggests that the most important deviations from standard GR will be evidenced at an early epoch only. 
This has an important consequence in what concerns the testability of the theory, because the early history of the Universe is very tightly constrained by several independent sources of evidence.

In order to examine more in detail the specific phenomenology, we have used dynamical system techniques. In the case of dust we have found that the resulting model is basically the one already analyzed in \cite{SES}, with the difference that now the sign and the strength of the matter couplings become free parameters or the theory. However, in the end this fact does not have any influences on the number of the fixed points nor on the solutions associated to them, but only on the set of initial conditions that lead to a certain behavior.

The case of radiation is more complex because of the additional terms that appear in the cosmological equations. In phase space such terms are associated to additional fixed points ($\mathcal{A}$, $\mathcal{C}-\mathcal{G}$). If all variables are negative, one can obtain the richest possible behavior for the corresponding cosmology, i.e., orbits that are able to ``touch'' the entire set of fixed points (albeit they are all unstable).  The presence of such fixed points and, in particular, the existence of the solutions associated to them, constitutes one of the main results of the present paper. In fact these fixed points allow us to draw quantitative conclusions on some of the best studied phenomena of the early Universe, as nucleosynthesis or cosmic microwave background (CMB) physics.

It is natural to ask if one can choose the above mentioned constants is such a way that the models develop and early-time accelerated expansion phase which might work as inflation. The answer to that is, unfortunately, negative, for the time being, and the associated dynamical system gives a clear way to see it. In fact, one can prove that, given a generalized Friedman equation of the form
\begin{equation} \label{}
H^2= \frac{\alpha_1}{a^{m_1}}+\frac{\alpha_2}{a^{m_2}}+... +\frac{\alpha_i}{a^{m_i}},
\end{equation}
only terms in which $0<m_i<2$ can generate fixed points which may represent accelerated expansion. This is clearly not the case for any of the terms appearing in \rf{FriedNoDB2} or in \rf{FriedRad}. What means that, in the phase space, there is no fixed point which could correspond to an inflationary era. However, the situation is not as bad as it might seem at first sight. It has been already suggested that, within the HL framework, other mechanisms might substitute an inflationary phase  \cite{Kcosmo}. Our results seem to support this last, alternative scenario instead of the classical one. Therefore, in spite of the degrees of freedom added by the additional matter couplings it is very difficult to reproduce any accelerated expansion and the only way in which one can achieve a dark energy era is to introduce the cosmological constant ``by hand" in the HL potential. This appears as a clear limitation of the theory since it does not seems to provided a natural solution to the quantum gravity phenomenon of dark energy\footnote{It is worth noting that the results above depend critically on the form of (\ref{couplingRho}-\ref{couplingP}). In our calculation we have chosen a relatively simple structure for the coupling terms, but, in principle one could consider less obvious combinations. In fact one could even reverse engineer these coupling to try to obtain a more desirable behavior. The resulting terms, however, do not have a simple or elegant structure and, for this reason, they have not been considered in the text. }.

Anyway, the important issue to be remarked is that, thanks to a consistent introduction of the matter field, the HL theory in this new form finally becomes directly testable against some of the most accurate data we posses at present. And these data hold the promise to be able to verify directly its validity.

\section*{Aknowledgments}

This work was partially funded by Ministerio de Educaci\'on y Ciencia, Spain, projects
CICYT-FEDER-FPA2005-02211, SGR2005-00916, UniverseNet (MRTN-CT-2006-035863), AP2006-03102, FIS2006-02842 and AGAUR, contract 2009SGR-994 and grant 2010BE-100058. SC was funded by Generalitat de Catalunya through the Beatriu de Pin\'{o}s contract 2007BP-B1 00136. EE's research was performed in part while on leave at Department of Physics and Astronomy, Dartmouth College, 6127 Wilder Laboratory, Hanover, NH 03755, USA.

\end{document}